\documentclass[prd,aps,floats,preprintnumbers,preprint]{revtex4}

\usepackage{graphicx, epsfig}
 
\newcommand{\lcdm}{$\Lambda$CDM}

\newcommand{\be}{\begin{equation}}
\newcommand{\ee}{\end{equation}}
\newcommand{\bea}{\begin{eqnarray}}
\newcommand{\eea}{\end{eqnarray}}

\newcommand{\etal}{{\it et al.}}

\begin{document}


\setlength{\unitlength}{1mm}

\title{Redshift spherical shell energy in isotropic Universes}

\author{Antonio Enea Romano}

\affiliation{Department of Physics, University of Wisconsin, Madison, WI 53706, USA 
}

\affiliation{
IAC , La Laguna, Tenerife, Spain
}

\begin{abstract}
We introduce the redshift spherical shell energy (RSSE), which can be used to test in the redshift space the radial inhomogeneity of an isotropic universe, providing additional constraints for LTB models, and a more general test of cosmic homogeneity.

\end{abstract}

\maketitle
\section{Introduction}

One of the main assumption of the standard cosmological model is spatial homogeneity, but as observed by \cite{Celerier:2006hu}, this is more a philosophical principle than an experimental direct evidence.
In fact homogeneity cannot be tested experimentally on the entire region of locally observable Universe, since  comparing densities ad different distances from us implies comparing different times as well.
The only direct signature of spatial homogeneity we can observe is hence homogeneity of concentric spherical shells, approximatively corresponding to the same time range.

Making some assumptions on the time scale on which the evolution of the particular class of astrophysical objects we decide to observe to test the homogeneity can be neglected, we could treat spherical shells of appropriate height $\Delta Z(z)$ as effective homogeneity probes.
 
Previous analyses aimed to test homogeneity within one of such spherical shells \cite{Hogg:2004vw}, concluding that the inhomogeneity of disjoint regions were within the limits predicted by \lcdm \,  models, even if as observed by the authors, the results had a strong dependence on the K-correction.
In another similar study \cite{scaramella98a} a sphere was used, showing that the galaxies number density had fractal dimension three for FRW models, when the radius is sufficiently large. 
Assuming that the statistical analysis and the K-correction were correct, these cannot be considered proves of homogeneity of the entire space but only of the spherical shells, i.e. of isotropy of the universe at the distance and time corresponding to the redshift range considered.

In order to study the homogeneity of the entire space a similar analysis should be performed for shells at different distances, but in this case the evolution time scale would play a very important role, and it may be difficult to compare shells at very different redshifts, since they correspond to very different  times and consequently different stages of astrophysical evolution. This will have a very strong impact on their luminosity, and consequently on their number counting.

Another very important evidence of cosmic isotropy comes from the cosmic microwave background radiation (CMB), which accurate observations have shown to be highly isotropic \cite{Spergel:2006hy,WMAP2003}.
Following a similar argument to the one above, CMB isotropy does not imply the homogeneity of the entire Universe at the time of last scattering $t_{ls}$, but it can only give information about a spatial spherical shell corresponding to the time interval over which we have been observing it.
At different times we are in fact observing different last scattering surfaces, corresponding to the region of space of the Universe at $t_{ls}$ which becomes visible at the time of observation. 
If the Universe is radially inhomogeneous, waiting for a sufficiently large time corresponding to the radial spatial scale at which radial inhomogeneity sets in, we could observe a different anisotropy spectrum or last scattering temperature $T_{LS}$.

Despite  of this intrinsic observational difficulty in testing spatial homogeneity, we can still try to use available data to test different cosmological models, and in the same way high redshift luminosity distance observation $D_L(z)$ have been shown \cite{Chung:2006xh,Alnes:2005rw} to be consistent with both a FRW model with cosmological constant and LTB models without dark energy, we could  find an appropriate inhomogeneous LTB model in agreement with the observed "radial spherical shell energy" (RSSE) $\rho_{SS}(z)$ .

\section{Lemaitre-Tolman-Bondi (LTB) Solution\label{ltb}}
Isotropy assumption, which is well supported by different experimental evidence such as CMB radiation for example, leads to the 
Lemaitre-Tolman-Bondi metric  \cite{Lemaitre:1933qe,Tolman:1934za,Bondi:1947av} :
\begin{eqnarray}
\label{eq1} %
ds^2 = -dt^2  + \frac{\left(R,_{r}\right)^2 dr^2}{1 + 2\,E(r)}+R^2
d\Omega^2 \, ,
\end{eqnarray}
where $R$ is a function of the time coordinate $t$ and the radial
coordinate $r$, $E(r)$ is an arbitrary function of $r$, and
$R,_{r}$ denotes the partial derivative of $R$ with respect to
$r$.

Einstein's equations give:
\begin{eqnarray}
\label{eq2} \left({\frac{\dot{R}}{R}}\right)^2&=&\frac{2
E(r)}{R^2}+\frac{2M(r)}{R^3} \, , \\
\label{eq3} \rho(t,r)&=&\frac{M,_{r}}{R^2 R,_{r}} \, ,
\end{eqnarray}
with $M(r)$ being an arbitrary function of $r$ and the dot
denoting the partial derivative with respect to $t$. The solution
of Eq.\ (\ref{eq2}) can be written parametrically by using a
variable $\eta=\int dt/R \,$, as follows.
\begin{eqnarray}
\label{eq4} \tilde{R}(\eta ,r) &=& \frac{M(r)}{- 2 E(r)}
     \left[ 1 - \cos \left(\sqrt{-2 E(r)} \eta \right) \right] \, ,\\
\label{eq5} t(\eta ,r) &=& \frac{M(r)}{- 2 E(r)}
     \left[ \eta -\frac{1}{\sqrt{-2 E(r)} } \sin \left(\sqrt{-2 E(r)}
     \eta \right) \right] + t_{b}(r) \, ,
\end{eqnarray}
where  $\tilde{R}$ has been introduced to make clear the distinction between the two functions $R(t,r)$ and $\tilde{R}(\eta,r)$ which are trivially related by 

\begin{equation}
R(t,r)=\tilde{R}(\eta(t,r),r)
\label{Rtilde}
\end{equation}

and $t_{b}(r)$ is another arbitrary function of $r$, called bang function, which corresponds to the fact that big-bang/crunches happen at different times in this space, since the scale factor is not the same everywhere as in a FRW model.

We can introduce the following variables
\begin{equation}
 a(t,r)=\frac{R(t,r)}{r},\quad k(r)=-\frac{2E(r)}{r^2},\quad
  \rho_0(r)=\frac{6M(r)}{r^3} \, ,
\end{equation}
so that  Eq.\ (\ref{eq1}) and the Einstein equations
(\ref{eq2}) and (\ref{eq3}) can be written in a form which is more similar to FRW models:

\begin{equation}
\label{eq6} ds^2 =
-dt^2+a^2\left[\left(1+\frac{a,_{r}r}{a}\right)^2
    \frac{dr^2}{1-k(r)r^2}+r^2d\Omega_2^2\right] \, ,
\end{equation}
\begin{eqnarray}
\label{eq7} %
\left(\frac{\dot{a}}{a}\right)^2 &=&
-\frac{k(r)}{a^2}+\frac{\rho_0(r)}{3a^3} \, ,\\
\label{eq:LTB rho 2} %
\rho(t,r) &=& \frac{(\rho_0 r^3)_{, r}}{6 a^2 r^2 (ar)_{, r}} \, .
\end{eqnarray}
The solution in Eqs.\ (\ref{eq4}) and (\ref{eq5}) can now be written as
\begin{eqnarray}
\label{LTB soln2 R} \tilde{a}(\tilde{\eta},r) &=& \frac{\rho_0(r)}{6k(r)}
     \left[ 1 - \cos \left( \sqrt{k(r)} \, \tilde{\eta} \right) \right] \, ,\\
\label{LTB soln2 t} t(\tilde{\eta},r) &=& \frac{\rho_0(r)}{6k(r)}
     \left[ \tilde{\eta} -\frac{1}{\sqrt{k(r)}} \sin
     \left(\sqrt{k(r)} \, \tilde{\eta} \right) \right] + t_{b}(r) \, ,
\end{eqnarray}
where $\tilde{\eta} \equiv \eta r = \int dt/a \,$.
 
Eqns. (\ref{eq7},\ref{eq:LTB rho 2}) clearly shows that LTB solution reduces to FRW when the $k(r)$ and $\rho(r)$ are constants.

The geodesic equations expressed in terms of the redshift can be written as 

\cite{Celerier:1999hp,Kristian:1965sz}

\begin{equation}
\frac{dr}{dz}=\frac{\sqrt{1+2E(r(z))}}{(1+z)\partial_{t}\partial_{r}R(t(z),r(z))}\label{eq:lumindist2}\end{equation}
\begin{equation}
\frac{dt}{dz}=\frac{-|\partial_{r}R(t(z),r(z))|}{(1+z)\partial_{t}\partial_{r}R(t(z),r(z))},\label{eq:lumindist3}\end{equation}
where $t(z)$ and $r(z)$ physically represent the geodesic of the photon coming to us (located at $r=0,z=0$) starting from the radial distance of our horizon.

For a flat matter dominated FRW model, we have for example:

\bea
\{k(r) &= &0,\rho(r)=const,t_b(r)=0\} \\
r(z) & = & \frac{2}{H_0}[1-\frac{1}{\sqrt{1+z}}] \\
t(z) & = & \frac{2}{3H_0}(1+z)^{-3/2}.
\eea

\section{Radial homogeneity}

Distance estimations based on redshift measurement assume a cosmological model, and are therefore model dependent.
Tests of cosmological models should consequently be based on redshift space data, in order to clearly distinguish between the cosmological observables and the model.
Previous homogeneity tests based on the analysis of galaxy catalogs have been performed according to different methods   \cite{Hogg:2004vw}, supporting homogeneity on a sufficiently large scales.
The observational data were analyzed assuming implicitly homogeneity by using the luminosity distance relation (LDR) $D_L(z)^{FRW}$ corresponding to a homogeneous and isotropic universe, described by a FRW metric.
Since distance measurements depend on the cosmological model through $r(z)$, fitting observed data to a homogeneous FRW model to prove homogeneity is not necessarily a proof of a spatial homogeneity, in the same way high redshift luminosity distance supernovae data can be fitted by both a FRW universe with positive cosmological constant and   inhomogeneous LTB models without invoking dark energy.
For example our Universe could be inhomogeneous on large scale, but using $r_{FRW}(z)$ to find the comoving coordinate of an observed astrophysical object we could obtain a measure of the position which is not the true physical one obtained using $r_{LTB}(z)$, corresponding to the LTB metric which correctly describes an inhomogeneous and isotropic space.
Since the only real direct observable is the red-shift, it is more convenient to define homogeneity in terms of z.

Assuming the evolution of the astrophysical objects we are studying can be neglected on a time scale $\Delta t$ during which their density and luminosity can therefore be considered constant, we define the redshift spherical shell height $\Delta Z $ as:

\bea
\label{dz1}
\Delta t & = &t(z)-t(z+\Delta Z) \\
\label{dz2}
\Delta Z(z) & = & t^{-1}[t(z)-\Delta t]-z
\eea

which show how it depends on the cosmological model through $t(z)$. 
 
Using the spatial volume element for LTB models  
\begin{equation} 
\label{eq:dvr}
dV(r,t)=dv(r,t)dr %
= 4 \pi \frac{R^2 R,_{r}}{\sqrt{1+2E(r)}} \, dr \,.
\end{equation}

we can now define the "redshift spherical shell energy" (RSSE) $\rho_{SS}(z)$ as:

\begin{figure}[h]
	\begin{center}
		\includegraphics[height=15cm,width=10cm,angle=-90]{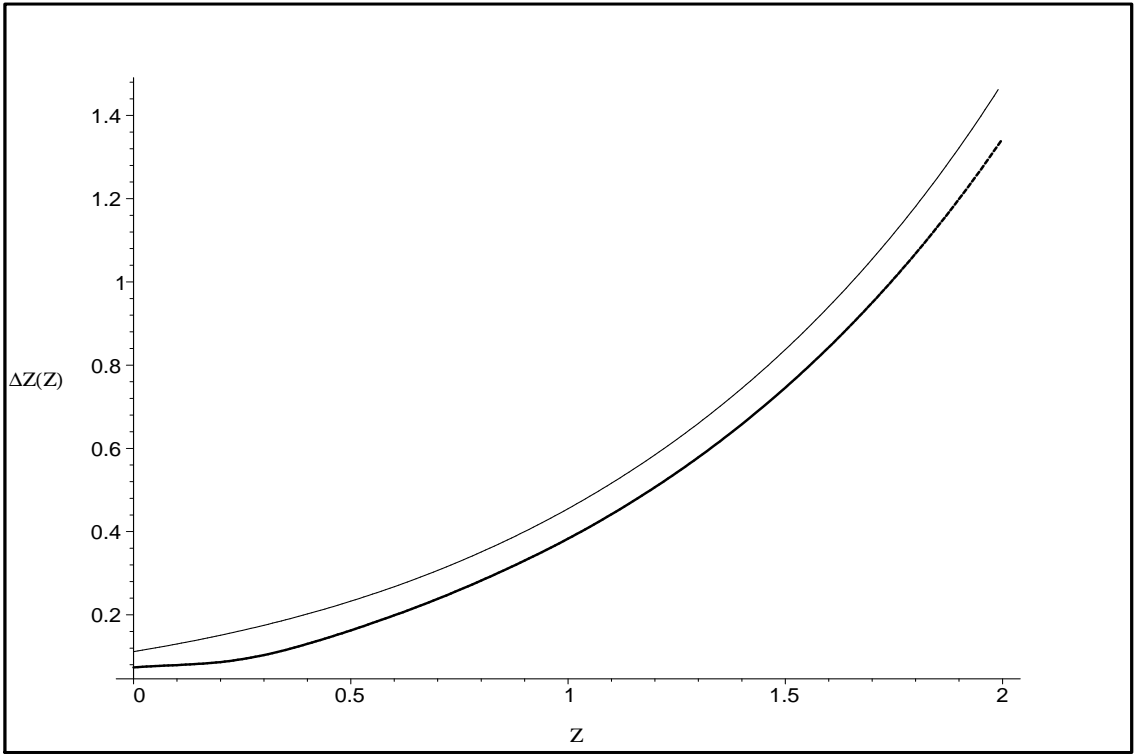}
	\end{center}
		\caption{$\Delta Z(z)$ is plotted as a function of the redshift for the LTB model considered (thick line) and for a FRW flat Universe with $\Omega_{\Lambda}=0.7,\Omega_M=0.3$ (thin line).}
	\label{fig:dzlm}
\end{figure}

%

\begin{figure}[h]
	\begin{center}
		\includegraphics[height=15cm,width=10cm,angle=-90]{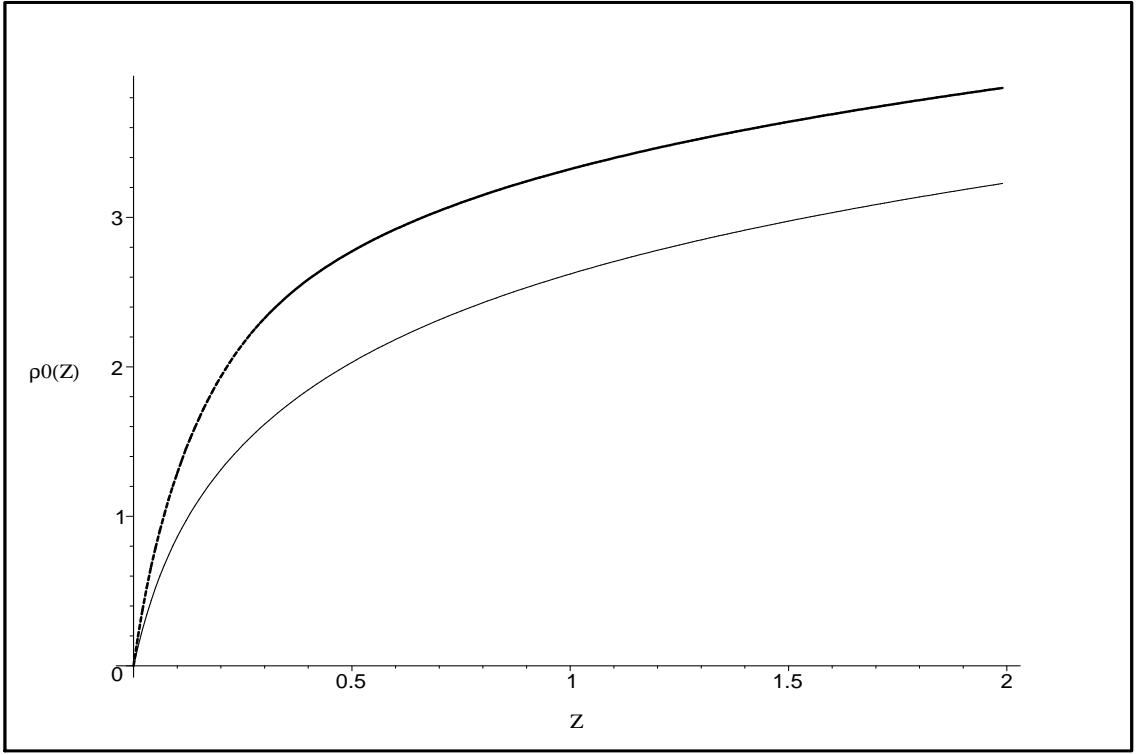}
	\end{center}
		\caption{ ${\rho_0(z)=Log_{10}[\rho_{SS}(z)/\rho_{SS}(0)]}$ is plotted as a function of the redshift for the LTB model considered (thick line) and for a FRW flat Universe with $\Omega_{\Lambda}=0.7,\Omega_M=0.3$ (thin line).}
	\label{fig:mz0LTB}
\end{figure}


\begin{eqnarray}
\label{rhoav2}	
\rho_{SS}(z) & = & 
\int\limits_{z}^{z+\Delta Z(z)}{\rho(z')dV(z')} =  \int\limits_{z}^{z+\Delta Z(z)}{4 \pi \frac{M_{,r}(r(z))}{\sqrt{1+2E(r(z))}}\frac{dr}{dz}dz} \\
\rho(z) & = & \rho(r(z),t(z)) \\
dV(z) & = &  dV(r(z),t(z)) =  dv(r(z),t(z))\frac{dr}{dz}dz
\end{eqnarray}



The RSSE can be considered the continuous equivalent of the galaxies number used in homogeneity tests, assuming their number density is representative of the total energy density $\rho(t,r)$.


\bea
\label{drz}
\label{mzr}
\rho_{SS}(z) & = &  \int\limits_{r(z)}^{r(z+\Delta Z(z))}{4\pi\frac{M_{,r}(r)}{\sqrt{1+2E(r)}}dr} .
\eea

The value of $\Delta t$ depends on the particular type of astrophysical objects considered, and requires a good understanding of their evolution, but this is beyond the scope of the present article.

In order to provide an example comparable with previous homogeneity studies, we will use the value implied by the redshift range $0.2<z<0.35$ studied by \cite{Hogg:2004vw}, assuming a flat FRW model with \{$\Omega_{M}=0.3,\Omega_{\Lambda}=0.7$\}:

\bea
\label{tfrw}
t(z)_{FRW} & = & H_0^{-1}\int\limits_{0}^{(1+z)^{-1}}{\frac{dx}{\left[\Omega_M x^{-1}+\Omega_{\Lambda}x^2\right]^{1/2}}} \\
\Delta t & = & t_{FRW}(0.2)-t_{FRW}(0.35)\approx1.4 Gyr \\
\Delta Z_{FRW}(0.2) & = & 0.35-0.2=0.15
\eea


In the case of a FRW Universe we get the simplified expression :
\bea
r(z)_{FRW} & = & H_0^{-1}\int\limits_{(1+z)^{-1}}^{1}{\frac{dx}{\left[\Omega_M x^{1}+\Omega_{\Lambda}x^4\right]^{1/2}}} \\
\rho_{SS}(z)_{FRW} & = & \int\limits_{z}^{z+\Delta Z(z)} \frac{\rho_0}{a(z)^3}a(z)^3 4\pi r(z)^2 d(r(z)) = \frac{4\pi\rho_0}{3}[r_{FRW}(z+\Delta Z(z))^3-r_{FRW}(z)^3] \\
\rho_{SS}(0)_{FRW} & \propto  & r_{FRW}(\Delta Z(0))^3
\eea

which shows that, as expected, $\rho_{SS}(0)$ scales as the third power of the comoving distance $r_{FRW}$, corresponding to the homogeneity condition tested by \cite{scaramella98a}.




In order to give an example of RSSE for different models, we considered two models  which both fit successfully the observed high redshift supernovae luminosity distance $D_L(z)$, a flat FRW model with $\{\Omega_{\Lambda}=0.7,\Omega_M=0.3\}$ and the LTB model proposed by \cite{Alnes:2005rw} corresponding to:
\begin{equation}
E(r)=\frac{1}{2}H_{\perp,0}^{2}r^{2}(\beta_{0}-\frac{\Delta\beta}{2}[1-\tanh\frac{r-r_{0}}{2\Delta r}])\label{eq:eansatz}\end{equation}
\begin{equation}
M(r)=\frac{1}{2}H_{\perp,0}^{2}r^{3}(\alpha_{0}-\frac{\Delta\alpha}{2}[1-\tanh\frac{r-r_{0}}{2\Delta r}])\label{eq:mansatz}\end{equation}
\begin{equation}
R(t_{rec},r)=a_{rec}r\label{eq:rinitcondagain},\end{equation}
with\begin{equation}
\{\alpha_{0}=1,\beta_{0}=0,\Delta\beta=-\Delta\alpha=-0.9,\Delta r=0.4r_{0},r_{0}\approx\frac{1}{5H_{0}},H_{\perp,0}\approx H_{0}\},\label{eq:params}\end{equation}
where $H_{0}\approx50\textrm{km/s/Mpc}$ and $a_{rec}\sim10^{-3}$
is the effective scale factor at recombination.

For both models we used the same $\Delta t$ calculated in eq.(\ref{tfrw}), but we would like to emphasize the fact that this choice has not any particular physical reason, but is simply the value implicitly assumed by \cite{Hogg:2004vw} in performing their homogeneity test. A different class of astrophysical objects, or different models of their evolution, would correspond to a different choice of $\Delta t$, but our purpose here is only to provide a concrete example rather than to study the general problem of the determination of $\Delta t$.
  
As it is shown in fig.(\ref{fig:dzlm}) , $\Delta Z(z)$ has a quite different behavior for the two studied models, due to their different time redshift relation $t(z)$, but it is in both cases an increasing function of $z$.
We can see from fig.(\ref{fig:mz0LTB}) that there is a clear distinction between $\rho_{SS}(z)_{FRW}$ and $\rho_{SS}(z)_{LTB}$,  which could be used to exclude one of the two models even at relatively low redshift.
For an LTB model $\rho_{SS}(z)$ would be another observable such as $D_L(z)$ which could be used to find $M(r),E(r),t_b(r)$ with a differential method similar to the one used in \cite{Chung:2006xh}, allowing to reduce the degeneracy of the mapping.
In fact as shown in \cite{Chung:2006xh} , given $E(r),D_L(z),t_b(r)$ we can always find the corresponding $M(r)$, making the relation between $D_L(z)$ and LTB models not one-to-one.
Another important question to answer would be if the observed $\rho_{SS}(z)$ can be fitted by a homogeneous model or not, and if, as in the case of the luminosity distance $D_L(z)$, there can be also different inhomogeneous models compatible with it.
Data from galaxies catalogs such as  2DF or SDSS could be used to find $\rho_{SS}(z)$ and thus provide an additional observational constraint on inhomogeneous cosmological models.

\section{Relation of RSSE  to astrophysical observations}
The quantity we have introduced to describe the radial energy distribution of isotropic universes, could be used to distinguish between different models only if we can relate it to some astrophysical observable. A natural candidate, which has  been used in previous homogeneity tests, would be the galaxy number count  $n(z)$ , but its relation to $\rho_{SS}(z)$ can be  rather complicated, due to different factors which should be taken into consideration.
In general we can write :
$$n(z)=F(z)\rho_{SS}(z)$$   

where $F(z)$ is function of the redshift which depends on different effects such as K-correction, distance selection effect,  the bias between barionic and dark matter, and the mass light ratio relation.
The determination of the relation between the total energy density $\rho_{SS}(z)$ and $n(z)$ is a general complex astrophysical problem, which goes beyond the purpose of this paper, that was to introduce a quantity in the redshift space, RSSE, able to encode the effects of the dependency of the energy distribution on both the time $t(z)$ and space $r(z)$ redshift relations, but it should be addressed in detail in any future attempt to relate RSSE to observations.

It should be observed that in principle $F(z)$ could add further degeneracy to the problem of distinguishing between different cosmological models, since different RSSE could be associated to the same $n(z)$.

\section{Discussion}
Redshift spherical shell energy has been introduced to test different isotropic cosmological models, and it has been computed in the case of a flat FRW and an LTB model, both fitting the high redshift supernovae luminosity distance data.
Computing the RSSE from the available observational data could  provide some new constraints on LTB solutions as viable cosmological models compatible with observations, or could exclude them as alternatives to the standard homogeneous isotropic models. 

There are several important points which we have not examined in details, related to the calculation of RSSE.
Astrophysical evolution could in fact play a very important role when comparing shells at very different redshift, introducing a time dependency  which is not included in the cosmological model, and could make more difficult to distinguish between one model and another.

An alternative application of the proposed method could consist in the calculation of RSSE for shells centered at different points in space time.
It would also be important to asses if the same RSSE could correspond to different homogeneous and inhomogeneous models, a degeneracy which could be possible as it has been shown for other cosmological observables such as the luminosity distance for example.


\begin{acknowledgments}
I thank M.~N.~Celerier, D. H. Chung and Martin Lopez-Corredoira for their useful comments.

\end{acknowledgments}

\end{document}